\documentclass[pdflatex,sn-mathphys-num]{sn-jnl}

\usepackage{graphicx}%
\usepackage{multirow}%
\usepackage{amsmath,amssymb,amsfonts}%
\usepackage{amsthm}%
\usepackage{mathrsfs}%
\usepackage[title]{appendix}%
\usepackage{xcolor}%
\usepackage{textcomp}%
\usepackage{manyfoot}%
\usepackage{booktabs}%
\usepackage{algorithm}%
\usepackage{algorithmicx}%
\usepackage{algpseudocode}%
\usepackage{listings}%

\usepackage{comment}

\usepackage{amsmath}

\usepackage{physics}
\usepackage{subfig}
\usepackage{float}
\usepackage{xcolor}

\newcommand{\arcsinh}{\text{arcsinh}}
\newcommand{\be}{\begin{equation}}
\newcommand{\ee}{\end{equation}}
\newcommand{\bea}{\begin{eqnarray}}

\newcommand{\eea}{\end{eqnarray}}
\newcommand{\nn}{\nonumber}

\newcommand{\MP}{M_\text{P}}

\usepackage{amsmath}
\usepackage{array}
\usepackage{tensor}

\def\a{\alpha}
\def\b{\beta}
\def\c{\chi}
\def\d{\delta}

\def\eps{\epsilon}

\def\l{\lambda}
\def\m{\mu}
\def\n{\nu}

\def\tns{\tensor}
\def\cR{\mathcal{R}}

\theoremstyle{thmstyleone}%
\theoremstyle{thmstyletwo}%

\theoremstyle{thmstylethree}%

\newcommand{\pat}{\partial}

\renewcommand{\a}{\alpha}
\renewcommand{\b}{\beta}
\renewcommand{\c}{\gamma}
\renewcommand{\d}{\delta}
\renewcommand{\l}{\lambda}

\newcommand{\rmd}{\mathrm{d}}

\begin{document}

\title[Quasi-pole inflation in metric-affine gravity]{Quasi-pole inflation in metric-affine gravity}

\author[1]{\fnm{Antonio} \sur{Racioppi}}

\affil[1]{National Institute of Chemical Physics and Biophysics, R\"avala 10, 10143 Tallinn, Estonia}


\abstract{We propose a new mechanism for inflationary model building in the framework of metric-affine gravity. Such a mechanism involves an inflaton non-minimally coupled with the Holst invariant. If the non-minimal coupling function has a zero point and it is very steep at that same point, the corresponding inflaton kinetic function will feature a quasi-pole behaviour, implying a canonically normalized potential featuring an exponential plateau, regardless of the shape of the original inflaton potential. The inflationary predictions in such a region are equivalent to the ones of Starobinsky inflation.}




\maketitle

\section{Introduction}\label{intro}
The current paradigm for explaining the homogeneity and flatness of the Universe at large scales is cosmic inflation \cite{Starobinsky:1980te,Guth:1980zm,Linde:1981mu,Albrecht:1982wi}, i.e. an accelerated expansion during the very early Universe. Additionally, it also addresses the origin of the small inhomogeneities observed in the Cosmic Microwave Background radiation. In its simplest version, inflation is usually formulated by considering a scalar field, the inflaton, embedded in Einsteinian gravity. The near-exponential expansion of the Universe is induced by the inflaton energy density.

The latest combination of Planck, BICEP/Keck and BAO data~\cite{BICEP:2021xfz} has sensibly reduced the allowed parameters space, indicating as favoured realizations the Starobinsky model~\cite{Starobinsky:1980te} and Higgs-inflation~\cite{Bezrukov:2007ep}. Both models can be described by a scalar field non-minimally coupled to gravity  (e.g.~\cite{Galante:2014ifa,Jarv:2016sow} and references therein). 

However, when theories exhibit non-minimal couplings to gravity, the \emph{choice} of the dynamical degrees of freedom becomes extremely relevant. 
In the more customary metric gravity, the connection is set to be the Levi-Civita one and the only dynamical degree of freedom is the metric tensor.  On the other hand,  in metric-affine gravity (MAG), both the connection and the metric are dynamical variables and their corresponding equations of motion will establish the eventual relation between them. When the gravity action features only the term linear in the curvature scalar and no fermions, the two approaches lead to equivalent theories (e.g.~\cite{BeltranJimenez:2019esp,Rigouzzo:2023sbb} and refs. therein), otherwise the theories are completely different~\cite{BeltranJimenez:2019esp,Rigouzzo:2023sbb,Koivisto:2005yc,Bauer:2008zj} and lead to different phenomenological predictions, as recently studied in
 (e.g.~\cite{Racioppi:2017spw,Jarv:2017azx,Racioppi:2018zoy,Kannike:2018zwn,Racioppi:2019jsp,Jarv:2020qqm,Gialamas:2020snr,Racioppi:2021ynx,Racioppi:2021jai,Lillepalu:2022knx,Gialamas:2023flv,Piani:2023aof,Barker:2024ydb,Dioguardi:2021fmr,Racioppi:2022qxq,Dioguardi:2022oqu,Dioguardi:2023jwa,Kannike:2023kzt,TerenteDiaz:2023kgc,Marzo:2024pyn,Iosifidis:2025wrv,Bostan:2025zdt,Gialamas:2025kef,Dioguardi:2025vci,Dimopoulos:2025fuq,Bostan:2025vkt,Dioguardi:2025mpp,Karananas:2025xcv,Barker:2025xzd,Barker:2025rzd,Barker:2025fgo} and refs. therein). Moreover, MAG permits not only one, but two two-derivative curvature invariants: the usual Ricci-like scalar and the Holst invariant~\cite{Hojman:1980kv,Nelson:1980ph,Holst:1995pc}, which can be used to construct new models~(e.g.~\cite{Hecht:1996np,BeltranJimenez:2019hrm,Langvik:2020nrs,Rigouzzo:2022yan,Shaposhnikov:2020gts,Pradisi:2022nmh,Salvio:2022suk,Piani:2022gon,DiMarco:2023ncs,Gialamas:2022xtt,Gialamas:2024jeb,Gialamas:2024iyu,Racioppi:2024zva,Racioppi:2024pno,Gialamas:2024uar,Racioppi:2025pim,He:2024wqv,He:2025bli,He:2025fij,Katsoulas:2025srh} and refs. therein).

The scope of this article is to study a new mechanism in MAG, where the inflaton scalar is non-minimally coupled with the Holst invariant and the non-minimal coupling function exhibits a zero point and it is very steep at that same point. As we will see later, this kind of setup will induce a canonically normalized inflaton potential with an exponential plateau, regardless of the shape of the original potential.

This article is organized as follows. In Section~\ref{sec:model} we introduce the action for our inflationary model in metric-affine gravity and show how the exponential plateau is generated. In Section~\ref{sec:inflation} we present the inflationary predictions of our construction. Finally, in Section~\ref{sec:conclusions} we summarize our conclusions. In addition, in Appendix~\ref{appendix}, we show how our results can be extended also to models featuring a non-minimal coupling between the inflaton and the Ricci scalar.

\section{Generic model for exponential plateau} \label{sec:model}
Our starting point is the action describing a real scalar $\phi$, playing the role of the inflaton, embedded in metric-affine gravity and non-minimally coupled to the Holst-invariant
\be 
S = \int d^4x\sqrt{-g}\left[\frac{M_P^2}{2} \left( {\cal R}+\tilde f(\phi)\tilde{\cal R} \right)  -\frac{\partial_\mu \phi \, \partial^\mu \phi}{2} - V(\phi) \right], 
\label{eq:Sstart} 
\ee
where $M_P$ is the reduced Planck mass, $V(\phi)$ is the inflaton potential,  $\tilde f(\phi)$ is the non-minimal coupling function, ${\cal R}$ and $\tilde{\cal R}$  respectively, a scalar and pseudoscalar  contraction of the curvature (the latter also known as the Holst invariant~\cite{Hojman:1980kv,Nelson:1980ph,Holst:1995pc}),
\be 
{\cal R} \equiv g^{\nu\sigma}  \tensor{\cR}{^\mu_\nu_\mu_\sigma}, \qquad  
\tilde{\cal R} \equiv  g_{\a\m} \epsilon^{\mu\nu\rho\sigma} \tensor{\cR}{^\a_\nu_\rho_\sigma},\label{eq:RRpdef}
\ee
where $\epsilon^{\mu\nu\rho\sigma}$ is the totally antisymmetric Levi-Civita tensor\footnote{The Levi-Civita tensor is defined as $\eps_{\mu\nu\rho\sigma} =\sqrt{-g} \bar{\varepsilon}_{\mu\nu\rho\sigma}$, with $ \bar{\varepsilon}_{\mu\nu\rho\sigma}$ being the Levi-Civita symbol with $ \bar{\varepsilon}_{0123}=1$. Note that the components of $\bar{\varepsilon}^{\mu\nu\rho\sigma}$ are equal to the components of ${\rm sign}(g)\bar{\varepsilon}_{\mu\nu\rho\sigma} = -\bar{\varepsilon}_{\mu\nu\rho\sigma}$.}. $\tns{\cR}{^\mu_\nu_\rho_\sigma}$ is the curvature tensor associated with the connection $\tns{\Gamma}{^\mu_\sigma_\nu}$
\be 
\tns{\cR}{^\mu_\nu_\rho_\sigma} =\partial_\rho \tns{\Gamma}{^\mu_\sigma_\nu} - \partial_\sigma \tns{\Gamma}{^\mu_\rho_\nu} +\tns{\Gamma}{^\mu_\rho_\m}\tns{\Gamma}{^\m_\sigma_\nu} -\tns{\Gamma}{^\mu_\sigma_\m}\tns{\Gamma}{^\m_\rho_\nu}  \ . \label{eq:Riemann}
\ee

We do not consider any other term\footnote{For a more generic starting action featuring a non-minimal coupling with $\cR$, see Appendix \ref{appendix}.} in action \eqref{eq:Sstart} in order to keep the model as minimal as possible, considering terms that can only be generated from the curvature tensor (i.e. no Nieh-Yan term e.g. \cite{Langvik:2020nrs} and refs. therein), with only the massless graviton and the inflaton as physical degrees of freedom  (i.e. no $\tilde{\cal R}^2$ term (e.g. \cite{Salvio:2022suk} and refs. therein)) and without terms that feature more than two derivatives (i.e. no ${\cal R}^2$ like terms (e.g. \cite{Annala:2021zdt} and refs. therein)).

We recall that in MAG, the connection  $\Gamma_{~\mu\nu}^{\rho}$ is not imposed to be the Levi-Civita one, but it is derived by solving the corresponding equation of motion. We also remind that, if $\Gamma_{~\mu\nu}^{\rho}$ is the Levi-Civita connection,  $\tilde{\cal R}$ vanishes\footnote{The careful reader might notice that $\tilde{\cal R}$ actually vanishes for any  $\Gamma_{~\mu\sigma}^{\rho}$ so that $\Gamma_{~\mu\nu}^{\rho}=\Gamma_{~\nu\mu}^{\rho}$.} and ${\cal R}$ equals the Ricci scalar $R$ derived from the Levi-Civita connection. x
MAG theories generally involve nonzero torsion, $T^\c{}_{\a\b} \equiv 2 \Gamma^{\c}_{[\a\b]}\neq 0$, and nonzero non-metricity, $Q_{\l\m\n} \equiv \nabla_\l g_{\m\n} \neq 0$.
Moreover, \eqref{eq:Sstart} is dynamically equivalent to the Einstein-Cartan framework because $Q_{\l\m\n}$ can be set to zero thanks to a projective symmetry of the action. Such an equivalence would be lost in presence of other invariants directly built from $Q_{\l\m\n}$.

After some standard computations, the action~\eqref{eq:Sstart} can be cast in terms of Einsteinian gravity. For a detailed explanation of the computations we refer the reader to  \cite{Langvik:2020nrs,Rigouzzo:2022yan} (and refs. therein). In the following we just give the highlights. Using the aforementioned projective symmetry we set the non-metricity  $Q_{\l\m\n}=0$. Therefore the connection can be written as
\bea \label{eq:GammaMAG}
  \Gamma^\c_{\a\b} &=& \bar\Gamma^\c_{\a\b} + K^\c{}_{\a\b} \ ,
\eea
where $\bar\Gamma_{\a\b}^\c$ is the Levi--Civita connection constructed from the metric $g_{\a\b}$ and $K_{\a\b\c}$, is the contortion, defined in terms of the torsion as
\be \label{K}
 K_{\a\b\c} \equiv \frac{1}{2} ( T_{\a\b\c} + T_{\c\a\b} + T_{\b\a\c} ) \, .
\ee
It is convenient to define also the torsion vectors
\bea \label{eq:Tvec}
  T^\b \equiv g_{\a\c} T^{\a\b\c} \ , \qquad \tilde T^\a \equiv \frac{1}{6} \epsilon^{\a\b\c\d} T_{\b\c\d} \, ,
\eea
and the torsion scalar
\be \label{eq:T:scalar}
 T \equiv \frac{1}{4} T_{\a\b\c} T^{\a\b\c} - \frac{1}{2} T_{\a\b\c} T^{\c\a\b} - T_\a T^\a \, .
\ee
Using eqs. \eqref{eq:GammaMAG}-\eqref{eq:T:scalar}, we can write the Ricci scalar and the Holst invariant in eq. \eqref{eq:RRpdef} as
\bea \label{q:RtildeR:MAG}
  \cal R &=&  R  + T + 2\bar\nabla_\a  T^\a  \\
  \tilde {\cal R}  &=&
  - 6 \bar\nabla_\a \tilde T^\a 
  + \frac{1}{2} \epsilon^{\a\b\c\d} T_{\mu\a\b} T^\mu{}_{\c\d} 
\eea
where $R$ is the curvature scalar and $\bar\nabla_\a$ is the covariant derivative, both constructed with the Levi-Civita connection. Solving the equation of motion for $ \Gamma^\c_{\a\b}$, we obtain the solution for the torsion as
\be \label{eq:T:sol}
  T_{\a\b\c} = 2 t_1(\phi) g_{\a[\b} \pat_{\c]} \phi + t_2(\phi) \epsilon_{\a\b\c}{}^{\mu} \pat_\mu \phi \, ,
\ee
with
\be 
   t_1= 2 \frac{ M_P \tilde f(\phi) \tilde f(\phi)  ' }{1 + 4 \tilde f(\phi) ^2} \qquad  
  t_2 = 2 \, \frac{ \tilde f(\phi)   - M_P \tilde f(\phi) ' }{1 + 4 \tilde f(\phi) ^2} \, , \label{eq:t12}
\ee
where $'$ indicates the derivative with respect to the argument of the function.
Combining eqs. \eqref{eq:Tvec}-\eqref{eq:T:sol}, inserting the result into action \eqref{eq:Sstart} and dropping a boundary term, we can rewrite the action in terms of Einsteinian gravity as
\be
  \label{eq:actionG} S_E = \int\rmd^4 x \sqrt{-g} \bigg\{ \frac{\MP^2}{2}  R - \frac{1}{2} \left[ 1 + 6 t_1^2 - \frac{3}{2} t_2^2 - 6 t_2 \Big[ M_P \tilde f'-  2 t_1 \tilde f \Big] \right] \partial_\mu \phi \, \partial^\mu \phi   - V(\phi) \bigg\} \ .
\ee
Using $t_1$ and $t_2$ given in eq. \eqref{eq:t12}, we finally obtain
\be 
S_{\rm E} =\int d^4x\sqrt{-g}\left[\frac{\MP^2}{2} R -\frac{1}{2}\partial_\mu \chi \, \partial^\mu \chi - V(\chi) \right],
\label{eq:SE}
\ee 
where the canonically normalized scalar $\chi$ is defined by
\be
\left( \frac{d\chi}{d\phi} \right)^2= k(\phi) \, , \qquad   k(\phi) = 1+\frac{6 M_P^2 \left[{\tilde f(\phi)}'\right]^2}{ \left[ 1+4{\tilde f(\phi)}^2 \right]} \, .    \label{eq:k(phi)}
\ee
It is well known that when $k(\phi)$ features a pole (or just a pronounced peak) (e.g. \cite{Iosifidis:2025wrv} and refs. therein), it induces in $V(\chi)$ a flat region\footnote{In case of a very peaked $k(\phi)$, the flat region will be only local, appearing as an inflection point in $V(\chi)$ (e.g. \cite{Racioppi:2024pno,Racioppi:2024zva,He:2025bli} and refs. therein).} that might be suitable for inflation. Since the denominator in eq.~\eqref{eq:k(phi)} is strictly positive, the existence of a pole is excluded and the only available option is a local maximum with $k(\phi) \gg 1$. By looking at eq. \eqref{eq:k(phi)}, it is intuitive to guess that the requirement for a peaked $k(\phi)$ would be a very big numerator and a very small denominator in the fraction after the ``1+'' in eq. \eqref{eq:k(phi)}. The first is easily achieved with a very large $\tilde f(\phi)'$. However, this might also induce a large $\tilde f(\phi)$, implying then a large denominator that would counterbalance the large numerator, with the net effect of a non-peaked $k(\phi)$. The solution would be to keep \emph{under control} $\tilde f(\phi)$ in presence of a large first derivative. The best scenario seems to be the concurrence of the minimum value of the denominator and a very big  $\tilde f(\phi)'$. The denominator in  eq. \eqref{eq:k(phi)} is strictly positive and its minimal value is 1. Therefore, as we will see, we can induce a flat region in $V(\chi)$, if it exists a value $\phi_0$ so that
\bea
\tilde f (\phi_0) &=& 0 \, , \label{eq:zero} \\
 \tilde\xi = M_P \, |\tilde f(\phi_0)'|  &\gg& 1 \, , \label{eq:fp:big}
\eea
where $\tilde\xi$ is a dimensionless parameter\footnote{Note that $\tilde f$ is dimensionless, therefore  $M_P \, \tilde f'$ is dimensionless as well and eq. \eqref{eq:fp:big} is well defined.}. Assuming eqs. \eqref{eq:zero} and \eqref{eq:fp:big}, if $k(\phi)>>1$,  then  we can easily approximate the behaviour of $k(\phi)$ nearby the maximum by neglecting the ``$1+$" term before the fraction in~\eqref{eq:k(phi)}, obtaining
\be
  k(\phi)  \simeq \frac{6 M_P^2 \left[{\tilde f(\phi)}'\right]^2}{ \left[ 1+4{\tilde f(\phi)}^2 \right]} \, .
    \label{eq:k(phi):app}
\ee
Using such an approximation, \eqref{eq:k(phi)} can be integrated exactly, providing a solution nearby the peak as
\be
 \chi \simeq \pm \sqrt{\frac{3}{2}} M_P \left\{ \arcsinh\left[2 \tilde f(\phi) \right] \right\} + \chi_0 \, ,
   \label{eq:chi:app}
\ee
where the sign ambiguity coming from eq. \eqref{eq:k(phi)} does not carry any physical meaning and will be addressed later. $\chi_0$ parametrizes the freedom in choosing the origin of $\chi$ and it will be fixed later as well. 
Looking at eq. \eqref{eq:chi:app} as a function of $\tilde f$, it might appear that no flattening is induced because the $\arcsinh$ function is known to be a slow varying function of the argument. However, this is only true away from the origin, where instead a relative steep appears. This is precisely in agreement with the condition in eq. \eqref{eq:zero}. On the other hand, eq. \eqref{eq:zero} is not sufficient alone in order to generate a local flattening in $V(\chi)$, but needs to be supported by the condition in eq. \eqref{eq:fp:big}, as can be easily double checked by using the chain rule of derivatives and computing $|\chi'(\phi)|=|\chi'(\tilde f)||\tilde f'(\phi)|$ .

Assuming the flattening induced by the field redefinition \eqref{eq:chi:app}, thanks to the conditions \eqref{eq:zero} and \eqref{eq:fp:big}, a \emph{big} change in $\chi$ will correspond to \emph{small} change in $\phi$. Therefore
 it is reasonable to assume that most of the slow-roll dynamics will take place nearby $\phi_0$. In such a case, it is convenient to expand $\tilde f$ at the first order around $\phi_0$ obtaining
\bea
 \tilde f (\phi) &\simeq& \tilde f(\phi_0) + \tilde f'(\phi_0) (\phi - \phi_0) \nn\\
&=& s_0 \, \tilde\xi \, \frac{ \phi - \phi_0 }{M_P}  \, , \label{eq:ftilde:app}
\eea
where we have used eq. \eqref{eq:zero} and \eqref{eq:fp:big} and we have defined $s_0 = \text{sign}\left(\tilde f'(\phi_0)\right)$. By using eq. \eqref{eq:ftilde:app}, we can then rewrite eq. \eqref{eq:k(phi):app} as
\be 
k(\phi)  \simeq \frac{6 \tilde\xi^2}{1+4 \tilde\xi^2 \left( \frac{ \phi - \phi_0  }{M_P}\right)^2} \, . \label{eq:k:quasipole}
\ee
Eq. \eqref{eq:k:quasipole} features a quasi-pole behaviour (a regularized pole in the language of \cite{He:2025bli}), meaning that for $\tilde\xi \to \infty$, the ``1+'' in the denominator can be neglected and \eqref{eq:k:quasipole} can be approximated by the pole function
\be 
k(\phi)  \simeq \frac{3}{2 \left( \frac{ \phi - \phi_0  }{M_P}\right)^2} \, , \label{eq:k:pole}
\ee
as long as $\phi$ is close to but not exactly equal to $\phi_0$. Note that in such a case the dependence on $\tilde\xi$ factorizes out and the eq. \eqref{eq:k:pole} is nothing but a rearranged $\alpha$-attractors kinetic function \cite{Galante:2014ifa}. Hence already from here we could infer that the ultimate result will be Starobinsky inflation. Nevertheless, we consider it useful for the reader to proceed with the discussion  using the quasi-pole kinetic function in eq. \eqref{eq:k:quasipole} and apply the strong-coupling limit later. Using  eq. \eqref{eq:k:quasipole}, the field redefinition \eqref{eq:chi:app} becomes
\bea
 \chi &\simeq& \pm \sqrt{\frac{3}{2}} M_P \left\{ \arcsinh\left[2 \, s_0  \, \tilde\xi \, \frac{ \phi - \phi_0 }{M_P} \right] \right\} + \chi_0 \, , \nn\\
 &=& \pm  \, s_0  \, \sqrt{\frac{3}{2}} M_P \left\{ \arcsinh\left[2 \tilde\xi \, \frac{ \phi - \phi_0 }{M_P} \right] \right\} + \chi_0 \, ,
   \label{eq:chi:app:app}   
\eea
which can be inverted as
\be
 \phi \simeq \phi_0 \pm  \, s_0 \frac{M_P}{2 \tilde\xi} \sinh \left(\frac{\sqrt{\frac{2}{3}} \left(  \, \chi-\chi_0 \right)}{M_P}\right) \, .
   \label{eq:phi:app:app}
\ee
Analogously to eq. \eqref{eq:ftilde:app}, assuming again that most of slow-roll dynamics will take place nearby $\phi_0$, we can also expand at the first order $V(\phi)$, obtaining
\bea
 V (\phi) &\simeq&  V(\phi_0) +  V'(\phi_0) (\phi - \phi_0) \nn\\
    &=&  \lambda M_P^4 +  \lambda_1  M_P^3 (\phi - \phi_0) \, , \label{eq:V:app}
\eea
where we have defined $\lambda M_P^4 = V(\phi_0)>0$ and $\lambda_1  M_P^3 = V'(\phi_0)$, so that $\lambda$ and $\lambda_1$ are dimensionless parameters. Note that, while $\lambda$ needs to be positive (as it sets the energy scale for inflation), $\lambda_1$ can take any sign, positive or negative. Plugging eq. \eqref{eq:phi:app:app} into eq. \eqref{eq:V:app}, we obtain
\be
 V (\chi) \simeq \lambda  M_P^4 \pm  \, s_0 \frac{\lambda_1 M_P^4}{2 \tilde\xi} \sinh \left(\frac{\sqrt{\frac{2}{3}} \left(  \, \chi-\chi_0 \right)}{M_P}\right)  \, . \label{eq:V:app:sinh}
\ee
Then, it is convenient to choose
\be
 \chi_0 = \pm s_0 \sqrt{\frac{3}{2}} M_P \arcsinh\left(\frac{2 \lambda  \tilde\xi}{\lambda_1}\right) \, , \label{eq:chi0}
\ee
so that $V=0$ for $\chi=0$, obtaining
\be
 V (\chi) \simeq    \lambda \, M_P^4 \left\{1  \pm  \, s_0 \frac{1}{2 \gamma} \sinh \left[ \frac{\sqrt{\frac{2}{3}} \left( \chi \right) }{M_P} \mp s_0 \arcsinh\left( 2 \gamma \right)\right]\right\}   \, \, ,\label{eq:V:app:sinh:centered}
\ee
where we have introduced $\gamma = \frac{ \lambda}{\lambda_1} \tilde\xi$. Note that, because of $\lambda_1$, $\gamma$ can change sign as well.
Using the properties of the hyperbolic functions, we can rewrite the potential \eqref{eq:V:app:sinh:centered} in a less compact but more intuitive form:
\bea
 V(\chi) &\simeq& \lambda \, M_P^4 \left\{1 +
 \frac{\sqrt{1+ 4 \gamma{}^4}}{2 \gamma} 
\left( \pm  \, s_0 \right)
    \sinh \left(\frac{\sqrt{\frac{2}{3}} \chi }{M_P}\right) 
 -  \cosh \left(\frac{\sqrt{\frac{2}{3}} \chi }{M_P}\right) \right\} \, , \nn\\
 &=& \lambda \, M_P^4 \left\{1 +
 \frac{\sqrt{1+ 4 \gamma{}^4}}{2 |\gamma|} 
 %
%
    \sinh \left(\frac{\sqrt{\frac{2}{3}} \chi }{M_P}\right) 
 -  \cosh \left(\frac{\sqrt{\frac{2}{3}} \chi }{M_P}\right) \right\} \, ,
\label{eq:V:app:sinh:expanded} 
\eea
where, without loss of generality, the sign ambiguity coming from eq. \eqref{eq:k(phi)} has been solved in the last line so that the prefactor of the $\sinh$ is always positive.
As mentioned before, we are operating in $\tilde\xi \gg 1$ limit, or equivalently in the $|\gamma| \gg 1$ limit. 
 Therefore applying such a limit to eq. \eqref{eq:V:app:sinh:expanded} and keeping the lowest order correction in $\frac{1}{\gamma}$, we obtain
\be
 V (\chi) \simeq  \lambda M_P^4 \left( 1-e^{-\frac{\sqrt{\frac{2}{3}} \chi }{M_P}} + \frac{ e^{\frac{\sqrt{\frac{2}{3}} \chi }{M_P}}}{16 \gamma ^4} \right) \, , \label{eq:V:app:exp:gamma}
\ee
which is an exponential plateau corrected by a term suppressed by a $\gamma^4$ factor. Taking the limit $\gamma \to \infty$, we easily find the well known approximation of the Starobinsky potential:
\be
 V (\chi) \simeq  \lambda M_P^4 \left(1-e^{-\sqrt{\frac{2}{3}}\frac{\chi }{M_P}}\right)  \, . \label{eq:V:app:exp}
\ee
 Note that \eqref{eq:V:app:exp} depends only on the initial parameter $\lambda$, which sets the inflationary energy scale in $V(\phi_0)$, but it is independent on any other details regarding the potential $V(\phi)$ or the non-minimal coupling $\tilde f (\phi)$.
 
Before proceeding to the inflationary phenomenology, we would like to remark that the conditions \eqref{eq:zero} and \eqref{eq:fp:big} are actually easy to realize. For instance, given a continuous strictly positive function $g(\phi)$, we can construct the following
\be
 \tilde f(\phi) = \delta + \xi g(\phi)
\ee
with $\xi \gg 1$ and $\delta<0$. It is easy to prove that for any $\phi=\phi_0$ in the domain of $g$ we can find a $\delta$ so that the conditions are \eqref{eq:zero} and \eqref{eq:fp:big} satisfied. Moreover, using the symmetry $\tilde f \to - \tilde f$ (see eq. \eqref{eq:k(phi)}), the same argument can be extended to other sign configurations, as long as $\delta$ and $\xi g(\phi)$ keep opposite signs so that a $\phi_0$ exists. Finally, we remark that our procedure extends the pole-regularization procedure of \cite{He:2025bli} to a larger class of models, included the one shown in \cite{He:2025bli} (which can be expressed as our eq. \eqref{eq:V:app:exp:gamma} during slow-roll).

\section{Inflationary predictions}\label{sec:inflation}

In this section we discuss the inflationary predictions of the scenario described by eq. \eqref{eq:V:app:exp:gamma}. Using the slow-roll approximation, all the inflationary observables can be derived from the scalar potential and its derivatives. First of all we define the potential slow-roll parameters:
\bea
\epsilon_V  (\chi) &=& \frac{M_P^2}{2}\left(\frac{V'(\chi)}{V(\chi)}\right)^2 \, , \label{eq:epsilon}
\\
\eta_V  (\chi) &=& M_P^2 \frac{V''(\chi)}{V(\chi)} \, . \label{eq:eta}
\eea
The expansion of the Universe is evaluated in number of e-folds, which is given by
\be
N_e =  \frac{1}{M_P^2} \int_{\chi_{\textrm{end}}}^{\chi_N} {\rm d}\chi \, \frac{V(\chi)}{V'(\chi)} ,
\label{eq:Ne}
\ee
where the field value at the end of inflation is given by $\epsilon  (\chi_{\textrm{end}}) = 1 $, while the field value $\chi_N$ at the time a given scale left the horizon is given by the corresponding $N_e$. 
The tensor-to-scalar ratio $r$ and the scalar spectral index $n_\textrm{s}$ 
are:
\bea
r  &=& 16\epsilon_V  (\chi_N) \,  , \label{eq:r} \\
n_\textrm{s}  &=& 1+2\eta_V  (\chi_N)-6\epsilon_V  (\chi_N) \, .  \label{eq:ns} 
\eea
Finally, the amplitude of the scalar power spectrum is
\be
 A _\textrm{s} = \frac{1}{24 \pi^2 M_P^4}\frac{V(\chi_N)}{\epsilon_V  (\chi_N)}  \simeq 2.1 \times 10^{-9} \, ,
 \label{eq:As:th}
\ee
whose experimental constraint~\cite{Planck:2018jri} usually fixes the energy scale of inflation. Applying the equations above to the potential \eqref{eq:V:app:exp:gamma}, we  find
\bea
\chi_N & \approx & \sqrt{\frac{3}{2}} M_P \ln \left(\frac{2 N_e}{3}\right) \left(1+\frac{N_e^2}{108 \gamma^4 \ln \left(\frac{2 N_e}{3}\right)}\right) \, , \label{eq:chiN:app} \\
 r &\approx & \frac{12}{N_e^2}\left(1+ \frac{N_e^2}{27 {\gamma}^4} \right) \, , \label{eq:r:app} \\
 n_s &\approx & 1-\frac{2}{N_e}\left(1 - \frac{N_e^2}{27 {\gamma}^4} \right)  \, , \label{eq:ns:app}
\eea
\bea
 A_s &\approx & \frac{\lambda N_e^2 }{18 \pi } \left(1-\frac{N_e^2}{27 \gamma ^4}\right) \label{eq:As:app}   \, ,
\eea
where we have used the large $N_e$ approximation and kept the leading order correction in $1/\gamma$. Note that the \emph{smaller} $\gamma$, the larger the predictions for $r$ and $n_s$ with the respect to the Starobinsky ones. Moreover we can use the result in eq. \eqref{eq:chiN:app} and evaluate a rough  lower bound on $\gamma$. It is intuitive to check that moving from eq. \eqref{eq:V:app:exp:gamma} to \eqref{eq:V:app:exp} requires
\be
  \frac{ e^{\frac{\sqrt{\frac{2}{3}} \chi }{M_P}}}{16 \gamma ^4} \ll e^{-\frac{\sqrt{\frac{2}{3}} \chi }{M_P}} \label{eq:lower:bound:eq} \, .
\ee
By inserting eq. \eqref{eq:chiN:app} into \eqref{eq:lower:bound:eq} and keeping the leading order term we obtain
\be
 \gamma \gg \sqrt{N_e} \label{eq:lower:bound:gamma} \, ,
\ee
where a numerical prefactor smaller than one has been dropped from the right hand side of eq. \eqref{eq:lower:bound:gamma}, in order to keep the bound more readable.
We stress that, after a proper identification of $\gamma$, the results in \eqref{eq:chiN:app}-\eqref{eq:As:app} coincide with the ones in \cite{Racioppi:2024pno}, apart for the leading order values of $\chi_N$, which differ by $\sqrt\frac{3}{2} M_P \ln 2$. 
Such a result can be understood as follows. Apart the overall normalization factor, the potential in \cite{Racioppi:2024pno} is essentially  eq. \eqref{eq:V:app:sinh:expanded} squared. Indeed, the strong coupling limit of \cite{Racioppi:2024pno} corresponds to eq. \eqref{eq:V:app:exp} squared. Let us consider then a generalized version of the potential in eq. \eqref{eq:V:app:exp} 
\be
 V_m (\chi) =  \lambda M_P^4 \left(1-e^{-\sqrt{\frac{2}{3}}\frac{ \chi }{M_P}}\right)^m  \, , \label{eq:V:exp:m}
\ee
where the term inside parentheses in eq. \eqref{eq:V:app:exp} has been raised to the $m$-th power. Note that for $m=2$, the potential is exactly the Starobinsky one. During slow-roll $e^{-\frac{\sqrt{\frac{2}{3}} \chi }{M_P}} \ll 1$ (cf. eq. \eqref{eq:chi:app}), therefore we can approximate the potential in the inflationary regime as 
\be
 V_m (\chi) \simeq  \lambda M_P^4  \left(1-m \, e^{-\sqrt{\frac{2}{3}} \frac{\chi }{M_P}}\right) \, . \label{eq:V:exp:m:app}
\ee
Now it is convenient to perform the following redefinition 
\be
\chi=\bar\chi +\sqrt{\frac{3}{2}} \ln (m) M_P \, , \label{eq:chi:bar}
\ee
which leads to
\be
 V_m (\bar\chi) \simeq  \lambda  M_P^4 \left(1- \, e^{-\sqrt{\frac{2}{3}} \frac{\bar\chi }{M_P}}\right) \, , \label{eq:V:exp:m:app:bar}
\ee
which is exactly the same as eq. \eqref{eq:V:app:exp}. Focusing on the $m=2$ case in eq. \eqref{eq:chi:bar}, we easily obtain the difference between the $\chi_N$ values in eq. \eqref{eq:chiN:app} and in \cite{Racioppi:2024pno}. 

To conclude we stress that our results do not apply just to \cite{Racioppi:2024pno}, but they encompass a larger class of theories (fitting the requirements in eqs. \eqref{eq:zero} and \eqref{eq:fp:big}), like\footnote{To be precise, \cite{Shaposhnikov:2020gts,Racioppi:2024zva,Racioppi:2025pim} allow for the conditions \eqref{eq:zero}, \eqref{eq:fp:big}, but such a configuration was not explicitly investigated in the aforementioned references.} $\mathcal{\tilde R}^2$ models \cite{Gialamas:2022xtt,Salvio:2022suk,He:2024wqv,He:2025bli}, $\tilde\xi-$attractors-like configurations \cite{Langvik:2020nrs,Shaposhnikov:2020gts,Racioppi:2024pno,Gialamas:2024uar} and natural metric-affine inflationary models \cite{Racioppi:2024zva,Racioppi:2025pim}.

\section{Conclusions} \label{sec:conclusions}
We studied a new class of inflationary models in the framework of metric-affine gravity. Such a class exhibits the inflaton, $\phi$, non-minimally coupled with the Holst invariant $\tilde {\cal R}$ via a function, $\tilde f(\phi)$. If such a function becomes zero at a certain value $\phi=\phi_0$ (see eq. \eqref{eq:zero}) and it is very steep at that point (i.e. $M_P \tilde f'(\phi_0) \gg 1$, see eq. \eqref{eq:fp:big}), then the canonically normalized inflaton potential always features an exponential plateau, regardless of the shape of the original potential $V(\phi)$. The inflationary predictions in such a region are equivalent to the ones of Starobinsky inflation. Such a mechanism can be very attractive for inflationary model building, particularly if the Planck-ACT \emph{tension} \cite{AtacamaCosmologyTelescope:2025nti,Ferreira:2025lrd,Balkenhol:2025wms} is solved in favour of the first or if the reheating temperature is low enough so that $N_e \sim 70$ \cite{Drees:2025ngb,Zharov:2025zjg}.

\bmhead{Acknowledgements}

This work was supported by the Estonian Research Council grants PRG1055,  PRG1677, RVTT3, RVTT7, TARISTU24-TK10, TARISTU24-TK3 and the CoE program TK202 ``Foundations of the Universe". This article is based upon work from the COST Actions CosmoVerse CA21136 and BridgeQG CA23130 supported by COST (European Cooperation in Science and Technology).

\begin{appendices}
\section{General action} \label{appendix}
In this Appendix we study the more general setup where also a non-minimal coupling to the curvature scalar is considered\footnote{The following computation is quite standard and can be found elsewhere in the literature (e.g. \cite{Langvik:2020nrs}), however we report it also here for completeness.}. We start with the following action
\be 
S_{\rm J}= \int d^4x\sqrt{-\underline g}\left[\frac{M_P^2}{2} \left( f(\varphi){\cal \underline R}+h(\varphi)\underline{\cal \tilde R} \right)  -\frac{\partial_\mu \varphi \, \partial^\mu \varphi}{2} - U(\varphi) \right], 
\label{eq:Sstart:nmc} 
\ee
where $V(\varphi)$ the Jordan frame potential of the inflaton $\varphi$,  $f(\varphi)$ and $h(\varphi)$ are non-minimal coupling functions, 
$\underline{\cal R}$ and $\underline{\tilde{\cal R}}$ are defined as in eq. \eqref{eq:RRpdef} but in terms of the $\underline{g}{}_{\mu\nu}$ metric.
As mentioned before in Section \ref{sec:model}, we do not consider any other term in action~\eqref{eq:Sstart:nmc} in order to keep the model as minimal as possible.
Now, our goal is to transform the action \eqref{eq:Sstart:nmc} into action \eqref{eq:Sstart} via a Weyl transformation.
First of all, let us remind how a Weyl transformation acts on the curvature invariants in the metric-affine gravity. In this case the connection  $\Gamma_{~\mu\sigma}^{\rho}$ and the metric  $\underbar{g}{}_{\mu\nu}$ are  independent variables, therefore the Riemann tensor ${\cal R}_{~\mu\nu\sigma}^{\rho}$, being built from $\Gamma$ and its first derivatives, is invariant under any transformation involving only the metric. On the other hand, the Ricci scalar
and the Holst invariant (see eq. \eqref{eq:RRpdef})
are explicitly dependent on the metric. Therefore under a rescaling of the metric,
\be
  g_{\mu\nu} = \Omega^2 \underline{g}{}_{\mu\nu} \, , \label{eq:Weyl}
\ee
$\cal R$ and $\cal{\tilde R}$\, rescale inversely,
\be
   {\cal R} = \frac{\underline{\cal R}}{\Omega^2} \, , \qquad {\cal \tilde R} = \frac{\underline{\cal \tilde R}}{\Omega^2} \label{eq:R:scaling} \, .
\ee
Hence, by chooing $\Omega^2=f(\phi)$, we get the following action
\be 
S = \int d^4x\sqrt{-g}\left[\frac{M_P^2}{2} \left( {\cal R}
+ \frac{h(\varphi(\phi))}{f(\varphi(\phi))}\tilde{\cal R} \right)  
- \frac{\partial_\mu \phi \, \partial^\mu \phi}{2} 
- \frac{U(\varphi(\phi))}{f(\varphi(\phi))^2} \right], 
\label{eq:S:final} 
\ee
where $\phi$ is defined by solving
\be
\left( \frac{d\phi}{d\varphi} \right)^2= \frac{1}{f(\varphi)} \, .
\ee
It is now immediate to see that, using the redefinitions 
\be
 \tilde f (\phi) = \frac{h(\varphi(\phi))}{f(\varphi(\phi))} \,  \qquad V(\phi) = \frac{U(\varphi(\phi))}{f(\varphi(\phi))^2}
\ee
action \eqref{eq:S:final} is the same as \eqref{eq:Sstart} and the results obtained in Sections \ref{sec:model} and \ref{sec:inflation} can be easily extended to scenarios featuring a non-minimal coupling with the curvature scalar.

\end{appendices}


\bibliography{references}

\end{document}